\documentclass[graybox,vecarrow]{svmult}

\usepackage[utf8x]{inputenc}
\usepackage{mathptmx}       
\usepackage{helvet}         
\usepackage{courier}        
\usepackage{type1cm}        
\usepackage{makeidx}         
\usepackage{graphicx}        
\usepackage{epstopdf}
\usepackage{amssymb}
\usepackage{amsfonts}
\usepackage{amsbsy}
\usepackage{amsmath}

\usepackage{bm}

\usepackage{multicol}        
\usepackage[bottom]{footmisc}

\usepackage{hyperref}

\makeindex             

\newcommand{\eq}{\begin{equation}}
\newcommand{\eqx}{\end{equation}}
\newcommand{\eqs}{\begin{equation*}}
\newcommand{\eqsx}{\end{equation*}}
\newcommand{\eqn}{\begin{eqnarray}}
\newcommand{\eqnx}{\end{eqnarray}}
\newcommand{\alg}{\begin{align}}
\newcommand{\algx}{\end{align}}

\newcommand{\f}[2]{\frac{#1}{#2}}

\newcommand{\cor}[1]{\left\langle{#1}\right\rangle}

\newcommand{\lm}{\lambda}

\newcommand{\sg}{\sigma}
\newcommand{\dl}{\delta}
\newcommand{\Dl}{\Delta}
\newcommand{\al}{\alpha}
\newcommand{\bt}{\beta}

\newcommand{\gm}{\gamma}

\newcommand{\eps}{\varepsilon}
\newcommand{\qqqq}{\quad\quad\quad\quad}

\newcommand{\nn}{{\cal N}}

\newcommand{\RR}{{\mathbb{R}}}

\newcommand{\red}[1]{#1}
\newcommand{\green}[1]{#1}
\newcommand{\black}[1]{#1}
\newcommand{\rsq}{{\mathfrak R^2}}

\begin{document}

\title*{The dynamics of quark-gluon plasma and AdS/CFT}
\author{Romuald A. Janik}
\institute{Romuald A. Janik \at Institute of Physics,
Jagiellonian University\\
ul. Reymonta 4,
30-059 Krak\'ow,
Poland\\ \email{romuald@th.if.uj.edu.pl}}
%
%
\maketitle

\textsl{\noindent
Lectures at the 5th Aegean summer school, ``From gravity to thermal gauge theories: the AdS/CFT correspondence'', Adamas, Milos Island, Greece, September 21-26, 2009}
~\\
~\\~\\~\\

\abstract{
In these pedagogical lectures, we present the techniques of the AdS/CFT correspondence which can be applied to the study of real time dynamics of a strongly coupled plasma system. These methods are based on solving gravitational Einstein's equations on the string/gravity side of the AdS/CFT correspondence.
We illustrate these techniques with applications to the boost-invariant expansion of a plasma system. We emphasize the common underlying AdS/CFT description both in the large proper time regime where hydrodynamic dynamics dominates, and in the small proper time regime where the dynamics is far from equilibrium. These AdS/CFT methods provide a fascinating arena interrelating General Relativity phenomenae with strongly coupled gauge theory physics.
}

\section{Introduction}

The current experimental program of heavy-ion collisions at RHIC and the forthcoming experiments at LHC open an interesting window onto properties of QCD matter at high temperatures, where it appears in the guise of a new phase --- the quark-gluon plasma. At asymptotically high temperatures it should be a free gas of quarks and gluons, however, at the experimentally accessible energies there are strong indications that the quark-gluon plasma is indeed a strongly coupled system (see e.g. \cite{janreview}).

This poses numerous problems for its theoretical description, yet at the same time makes its study theoretically interesting. One can roughly  differentiate the physical properties of the quark-gluon plasma system into two broad categories --- static and dynamic (real-time) properties. 

The first of these, the static properties, typically involve the study of equilibrium thermodynamics, the entropy, energy density as a function of temperature and more generally properties which can be directly deduced from the Euclidean formulation of finite temperature gauge theory. In this case lattice QCD is an effective tool for accessing these properties in the nonperturbative, strongly coupled regime. It deals directly with QCD and yields quantitative results directly applicable for the QCD quark-gluon plasma.

The second class, the real time dynamic properties of strongly coupled plasma are much more difficult to access. They have to be formulated directly in Minkowski space and since lattice QCD methods are inherently Euclidean, it is very difficult to extrapolate numerical results to Minkowski signature. Moreover, it is exactly these kind of properties which are particularly relevant for the quark-gluon plasma produced in heavy-ion collisions. 

To this end let us recall schematically the basic stages of a heavy-ion collision. First the two ultrarelativistic nuclei collide and the plasma is produced in a state very far from equilibrium. Then in a relatively very short time, it becomes thermalized (or at least the pressure becomes isotropic with the residual anisotropy wholly due to flow). From that point on, hydrodynamic phenomenological models seem to describe the properties of the expanding plasma quite well 
\cite{janhydro}. In particular the plasma expands and cools, and when the temperature reaches the confinement/deconfinement phase transition one expects hadronization to occur.

It would be interesting to understand these various stages of the plasma dynamics directly from first principles. E.g. one would like to \emph{derive} the hydrodynamic behaviour from some theoretical framework and not only use it as a phenomenological model. But what is even more interesting is the understanding of the process of thermalization and what governs the short thermalization time necessary for the applicability of hydrodynamic models.

Unfortunately, for the case of QCD we lack appropriate theoretical methods which would be applicable to these kinds of problems at strong coupling. A possible route that one may take is to consider an analogous set of problems in a different theory for which appropriate real-time nonperturbative tools exist. 

The new method for studying nonperturbatively various gauge theories (although not directly QCD) is the AdS/CFT correspondence \cite{janadscft}, which translates dynamical problems in strongly coupled gauge theory into (usually) gravitational ones in higher number of dimensions. Its main advantage is that it works equally well in Minkowski as well as in Euclidean signature. In these lectures we will consider the AdS/CFT correspondence in its simplest original setting for the maximally supersymmetric $\nn=4$ Super-Yang-Mills theory. Of course, one has to be aware that the cost of switching the theory of interest from QCD to $\nn=4$ SYM is that we may most probably loose direct quantitative applicability of our results to realistic heavy-ion collisions. Moreover, there are certain marked differences between the theories which will cause a huge difference for certain physical phenomenae, at the same time being unimportant for other questions. We will discuss some of these points in these lectures.

Nevertheless, we would like to point out that currently we do not have 
\emph{any} gauge theory in which we would have a theoretical understanding of the issues described earlier. Therefore it is very interesting to study these issues for the case of $\nn=4$ SYM and use the results as a point of reference for analyzing the situation in QCD. Later, one could try to generalize these results to more complicated versions of the AdS/CFT correspondence for theories closer to QCD. In essence, this motivation is in line with the statement that {\it $\nn=4$ SYM is the harmonic oscillator of four dimensional gauge theories}. If one tries to develop some theoretical tools, one should better first apply them to the `harmonic oscillator'.

We have tried to make these lectures very pedagogical and self-contained. Our main emphasis in the presentation is to show how one can use the AdS/CFT correspondence as a tool even for far from equilibrium configurations without presupposing any kind of dynamics (which are in fact not known for nonlinear far from equilibrium systems). Therefore our presentation of hydrodynamics is 
subordinate to this goal, especially as a very general discussion focused on hydrodynamics \emph{per~se} is contained in the lectures by V. Hubeny at this school \cite{janhubenylect}.

The plan of these lectures is as follows. In section \ref{jans2}, we introduce the AdS/CFT correspondence, in section \ref{jans3} we compare some properties of plasma in the 
$\nn=4$ theory and in QCD. Then we proceed to  present the AdS/CFT setup specialized to the study of time-dependent dynamics of strongly coupled plasma. 
We then illustrate these methods in section \ref{jans5} by discussing two important examples -- the appearance from this setup of the standard planar AdS black hole, and a planar shock wave. In this section we also discuss some subtleties arising with different choices of coordinate systems which will be relevant later. Then in section \ref{jans6}, we introduce the main physical example of  a time-dependent plasma configuration --- the boost invariant flow. In section \ref{jans7}, we analyze its large proper time asymptotics and show how nonlinear perfect fluid dynamics arises from the AdS/CFT methods. In section \ref{jans8}, we show how one can see first corrections coming from shear viscosity, and in section~\ref{jans9}, for completeness, we will summarize briefly the current status of hydrodynamics in AdS/CFT. Then in section \ref{jans10}, we introduce physical situations, where the plasma dynamics is not describable by hydrodynamics, and finally, in section 
\ref{jans11}, we apply the AdS/CFT methods to study boost invariant flow in the far from equilibrium small proper time regime. We close the lectures with conclusions and an appendix with a short guide to the literature regarding work done on related topics which were not covered here.

\section{The AdS/CFT correspondence}
\label{jans2}
\index{The AdS/CFT correspondence}

The AdS/CFT correspondence \cite{janadscft}, in its original form states the equivalence 
of two apparently completely different theories: the $\nn=4$ supersymmetric
Yang-Mills theory (SYM) in four dimensions and type IIB superstrings in a ten-dimensional curved $AdS_5 \times S^5$ background. Since then, it has been generalized in various directions, extending it to a wider class of gauge theories, at the cost of making the dual string backgrounds more complicated. Throughout these lectures we will stay within the context of the AdS/CFT correspondence for $\nn=4$ SYM theory.

The reason why the AdS/CFT correspondence is so interesting is that the nonperturbative strong coupling regime of the $\nn=4$ gauge theory is mapped to the (semi-)classical strings or just (super)gravity which, in contrast to the gauge theory side, is at least theoretically tractable. Therefore one can use the AdS/CFT correspondence as a new method for accessing the very difficult nonperturbative gauge theory physics. 

The AdS/CFT correspondence is an equivalence, so in principle any state/pheno\-menon on the gauge theory side should have its direct counterpart on the string side and vice-versa. However one should keep in mind that the correspondence is an equivalence of gauge and \emph{string} theory, so the dual counterpart does not have to be in the well understood (super)gravity sector.
Fortunately, it will turn out that for the questions considered in these lectures namely the study of the dynamics of plasma expansion, the dual description will be purely gravitational.

Apart from the direct physical interest, the AdS/CFT correspondence is also theoretically very interesting as it translates various dynamical gauge theory questions into a geometrical language described by higher-dimensional General Relativity (GR). This leads to quite fascinating links between the two fields, providing a whole range of physically motivated interesting questions which could be addressed by GR methods. In the other direction, various notions introduced by the GR community like dynamical apparent horizons find their application and new interpretation on the gauge theory side.

\subsubsection*{Effective degrees of freedom at strong coupling}

As an illustration of the use of the AdS/CFT correspondence, and as a justification for the gravitational methods let us consider the question of finding effective degrees of freedom for strongly coupled $\nn=4$ SYM. By the AdS/CFT equivalence, it amounts to asking the same question for superstrings in $AdS_5 \times S^5$.

Let us first recall the case of closed strings in flat space. The string worldsheet  action is characterized by a dimensionfull parameter $\al'$ (related to the string tension). The various vibrational modes of the string correspond to particles (fields) with distinct masses
\eq
\label{jane.mflat}
m^2_n =\f{n^2}{\al'}
\eqx 
The massless modes correspond to the graviton (and its whole supergravity multiplet). There is also an infinite tower of massive modes. In the limit of 
$\al' \to 0$, the massive modes become very heavy and effectively decouple at fixed energies leaving as the governing dynamics just (super)gravity.

In the case of strings in $AdS_5 \times S^5$, the $\al'$ parameter becomes proportional to $1/\sqrt{\lm}$ where $\lm=g^2_{YM} N_c$ is the 't Hooft coupling of the dual $\nn=4$ gauge theory. The vibrational modes again split into a massless (super)graviton multiplet and a set of massive modes. The formula (\ref{jane.mflat}) is no longer exact, but the parametric behaviour with $\al'$ still holds. So in the strong coupling limit, those massive string modes become very heavy and effectively decouple leaving essentially supergravity modes as the effective degrees of freedom. Since the AdS/CFT correspondence postulates an equivalence with gauge theory, these should also correspond to the effective degrees of freedom of the gauge theory at strong coupling.

Once we lower the coupling, the massive modes become lighter and their effects will no longer be negligible. Initially, their effects may be absorbed into  
corrections to the gravitational Einstein-Hilbert action (so-called $\al'$ corrections), but at low coupling corresponding to the perturbative regime the spacetime description is not known.

Finally, let us note that the $\nn=4$ SYM theory is quite special in that it allows for such a clean separation between gravity modes and massive string modes. Presumably a dual description of real QCD (or even of large $N_c$ pure YM) would not have such a property.

\section{Why study $\nn=4$ plasma?}
\label{jans3}
\index{Quark-gluon plasma}
\index{QCD plasma vs. $\nn=4$ SYM plasma}

Since we will be using the AdS/CFT correspondence for $\nn=4$ SYM as a calculational tool for analyzing strongly coupled dynamics of gauge theory plasma, we will be essentially considering plasma in the supersymmetric $\nn=4$ gauge theory. This theory is completely different from QCD at zero temperature. It is supersymmetric, exactly conformal, does not have confinement. However once we turn on some nonzero temperature (or consider not the vacuum but some appropriate state), supersymmetry is broken and temperature (or energy density) introduces a scale. So qualitatively, we may expect to have similarities with QCD plasma in a certain window of temperatures where it is strongly coupled, approximately conformal and (by definition) deconfined.

However we have to keep in mind some definite differences w.r.t. QCD plasma. Firstly, the $\nn=4$ theory has no running coupling so, in contrast to QCD, even at very high tempeartures/energy densities the coupling may remain large. Secondly, the equation of state of the $\nn=4$ plasma is exactly conformal 
($E=3p$) which is only an approximation for a certain range of temperatures for QCD plasma. Indeed we know, from lattice QCD, that deviations from a conformal equation of state are important close to $T_c$.
In addition, we have other consequnces of conformality like that the bulk viscosity for the $\nn=4$ theory is exactly zero. Thirdly, for the $\nn=4$ theory (on Minkowski spacetime) there is no phase transition --- no analog of the confinement/deconfinement phase transition of QCD. Therefore as the plasma expands and cools, in the $\nn=4$ theory it will expand indefinitely, while in QCD it will cool down to the phase transition temparature and hadronize.

From the above discussion we see that the applicability of using $\nn=4$ plasma to model real world phenomena depends on the questions asked. It may give a good qualitative picture for the range of temperatures where we have similarities.
However, let us note that in this theory we may compute the dynamics from `first principles' (using the AdS/CFT correspondence). For QCD, unfortunately, we do not have any similar calculational technique, even numerical, which would enable us to study real-time dynamics of the strongly coupled quark-gluon plasma. Therefore it is interesting to build up results on strong coupling properties of $\nn=4$ plasma and use them as a point of departure for analyzing or describing QCD plasma.  Eventually, one might consider more realistic theories with AdS/CFT duals which are closer to QCD. In those cases generically the dual gravitational backgrounds are much more complicated so it is advantageous to start from the simplest setting for the $\nn=4$ SYM theory.

Another motivation for studying the dynamics of $\nn=4$ plasma is that the natural language of the AdS/CFT correspondence is quite new w.r.t. conventional gauge theory methods. So by studying relatively simple examples we may build up some new physical intuitions within this novel language. Also the interrelations with General Relativity physics are fascinating from the purely theoretical point of view. Last but not least, there may be some unexpected discoveries like the celebrated universality of the shear viscosity to entropy ratio $\eta/s$ \cite{janson}, which, at strong coupling, remains equal to $1/4\pi$ for \emph{any} theory with a gravitational dual \cite{januniv} (see the lectures by A. Starinets at this school \cite{janStarinetslect}).

\section{The AdS/CFT setup for studying real-time dynamics of plasma} 
\label{jans4}

Let us now briefly review the AdS/CFT correspondence on a more technical level, concentrating on the features relevant to the study of the time evolution of a plasma system.

The $S^5$ factor in the $AdS_5 \times S^5$ background is associated with a global 
$SO(6)=SU(4)$ symmetry of the $\nn=4$ theory.
In the following, we will only consider systems which do not break this symmetry, so the $S^5$ factor will be irrelevant and the whole dynamics will be concentrated in the $AdS_5$ factor.

The 5-dimensional Anti-de-Sitter spacetime $AdS_5$ can be given by the following metric
\eq
ds^2=\f{\eta_{\mu\nu} dx^\mu dx^\nu +dz^2}{z^2}
\eqx
with $z\geq 0$. $z=0$ is the boundary of $AdS_5$, while the region $z>0$ is often called `the bulk'. This choice of coordinates covers the Poincare patch of global $AdS_5$ and is relevant for the case when the dual gauge theory lives in $\RR^{1,3}$ Minkowski spacetime.
The above geometry can be understood to correspond to the gauge theory vacuum state. In particular gauge theory operators like the energy-momentum tensor 
$T_{\mu\nu}$ all have vanishing expectation values in this state
\eq
\cor{T_{\mu\nu}}=0
\eqx

Let us recall that the AdS/CFT correspondence states the equivalence with superstrings in $AdS_5$($\times S^5$). So, on the string side, we may excite any
normalizable mode, in particular we may excite gravitons. This will correspond to some states in $\nn=4$ SYM with $\cor{T_{\mu\nu}}\neq 0$. 
We expect a configuration of $\nn=4$ plasma to be a very complicated state which would correspond to exciting very many gravitons. Then it is better to interpret it instead as a change of the background:
\eq
\label{jane.ggen}
ds^2=g^{5D}_{\al\bt} dx^\al dx^\bt = \f{g_{\mu\nu}(x^\rho,z) dx^\mu dx^\nu +
dz^2}{z^2}
\eqx
with the metric coefficients being now generic functions of all the five coordinates.
We therefore seek to describe a plasma configuration in terms of the geometry 
$g_{\mu\nu}(x^\rho,z)$. Let us note that the above choice of the metric 
(\ref{jane.ggen}) is always possible after a suitable change of coordinates. Such coordinates, in which the metric has the form (\ref{jane.ggen}) are called Fefferman-Graham coordinates.
\index{Fefferman-Graham coordinates}

The geometry (\ref{jane.ggen}) cannot be, however, completely arbitrary. It must form a consistent background for strings, so it must satisfy 5-dimensional Einstein's equations with a negative cosmological constant\footnote{These equations are equivalent to the original ten-dimensional type IIB supergravity equations when we preserve full $SO(6)$ symmetry of $S^5$ and no other fields are turned on. The negative cosmological constant is a remnant of the RR 5-form under this dimensional reduction.}:
\eq
R_{\al\bt}-\f{1}{2}g^{5D}_{\al\bt} R - 6\, g^{5D}_{\al\bt}=0
\eqx
Furthermore, for a physical state in the gauge theory this geometry should not have a naked singularity. This turns out to be a crucial requirement with far reaching consequences for the resulting dynamics of the $\nn=4$ plasma, as we will see in the following.

\subsubsection*{The gravity $\longrightarrow$ $\cor{T_{\mu\nu}}$ dictionary}
\index{Holographic renormalization}

Once one has the geometry (\ref{jane.ggen}) corresponding to some plasma configuration, the key question is what is the energy momentum tensor of that
gauge theory system. The way to derive the answer, \emph{holographic renormalization}, has been explained in the lectures by K. Skenderis \cite{janskenderislect}. Here we just summarize the outcome derived in \cite{janskendorg}, which in the Fefferman-Graham coordinates defined by (\ref{jane.ggen}) takes a particularly simple form.

Suppose that the metric coefficients $g_{\mu\nu}(x^\rho,z)$ have the following Taylor expansion near the boundary\footnote{Here we always assume that the gauge theory lives in flat Minkowski space, hence the leading $\eta_{\mu\nu}$ and the absence of a $z^2$ term (see \cite{janskendorg} for details).}
\eq
g_{\mu\nu}(x^\rho,z)=\eta_{\mu\nu}+z^4 g^{(4)}_{\mu\nu}(x^\rho)+\ldots
\eqx
Then the expectation value of the energy-momentum tensor is
\eq
\label{jane.tmunuextract}
\cor{T_{\mu\nu}(x^\rho)} = \f{N_c^2}{2\pi^2} \cdot g^{(4)}_{\mu\nu}(x^\rho)
\eqx
The spacetime dependence of the energy-momentum tensor carries a lot of information about the dynamics of the plasma -- its energy density, momentum flow, stress tensor. Indeed, it is just this information which is exactly the direct outcome of hydrodynamic simulations of realistic heavy-ion collisions. Finally, let us emphasize, to avoid any chance of confusion, that the $T_{\mu\nu}$ is the energy-momentum tensor of the dual gauge theory. On the gravity side we are always dealing with \emph{vacuum} Einstein's equations.  

The construction outlined above leads to the following scenario of investigating a plasma system in strongly coupled $\nn=4$ SYM. One starts from some initial conditions for the 5-dimensional Einstein's equations. Then the geometry is evolved forward in time by solving Einstein's equations. Finally using the above formula~(\ref{jane.tmunuextract}), one extracts the $\cor{T_{\mu\nu}}$ of the corresponding plasma system. The details of the evolution of $T_{\mu\nu}$ are very interesting from the point of view of physics, especially in the far from equilibrium region, where very little is known about thermalization/isotropisation and transtion to a hydrodynamic expansion. We will follow this route in section \ref{jans11}.

\subsubsection*{The $\cor{T_{\mu\nu}}$ $\longrightarrow$ gravity dictionary}

It turns out to be very fruitful to ask also the opposite question. Suppose that we are given a certain spacetime profile of the energy momentum tensor $\cor{T_{\mu\nu}}$ -- how to construct the dual 5-dimensional geometry? The prescription is really just running the preceding recipe backwards. One has to solve Einstein's equations 
\eq
R_{\al\bt}-\f{1}{2}g^{5D}_{\al\bt} R - 6\, g^{5D}_{\al\bt}=0
\eqx
with the boundary condition 
\eq
\label{jane.bci}
g_{\mu\nu}(x^\rho,z)=\eta_{\mu\nu}+z^4 g^{(4)}_{\mu\nu}(x^\rho)+\ldots
\eqx
where $g^{(4)}_{\mu\nu}(x^\rho)$ is related to $\cor{T_{\mu\nu}(x^\rho)}$ through
\eq
\label{jane.bcii}
g^{(4)}_{\mu\nu}(x^\rho) = \f{2\pi^2}{N_c^2} \cor{T_{\mu\nu}(x^\rho)}
\eqx
It turns out that for a solution to exist, $g^{(4)}_{\mu\nu}(x^\rho)$ has to be traceless and conserved, which is of course physically expected for an energy-momentum tensor in a conformal theory. However here it is just an independent consequence of 5-dimensional Einstein's equations.

Once such a  $g^{(4)}_{\mu\nu}(x^\rho)$ is chosen, Einstein's equations determine uniquely the solution in the bulk (at least locally i.e. all higher coefficients of the Taylor expansion of $g_{\mu\nu}(x^\rho,z)$ around $z=0$ are uniquely determined).

In this way we see that for every energy momentum profile which does not violate the standard requirements of energy-momentum conservation and tracelessness we may construct a dual gravity background. However generically, such a geometry will be highly singular with naked singularities in the bulk. The requirement of the absence of naked singularities will very strongly constrain the admissible bulk geometries and hence also the possible spacetime profiles of the energy momentum tensor. This is a nontrivial constraint on the dynamics of the gauge theory as the spacetime profile includes of course the time evolution of $T_{\mu\nu}$. 

This line of reasoning was introduced in \cite{janUS1} as a way of determining the possible evolution of the energy momentum tensor. One first picks a family of profiles $T_{\mu\nu}(x^\rho)$, then one constructs for each of them a dual geometry by solving Einstein's equations with appropriate boundary conditions. Finally, one picks the allowed dynamics by requiring that the corresponding dual geometry is nonsingular. We will describe this procedure in detail in the first part of these lectures.

\section{Exact analytical examples}
\label{jans5}

In this section we will illustrate the AdS/CFT methods by analyzing two simple examples of plasma configurations for which the dual gravitational background can be computed in closed form \cite{janUS1}. Both of these examples have also a clear physical interpretation.

\subsection{A case study: static uniform plasma}

Let us first consider the simplest configuration of plasma, namely with a  uniform and static distribution of energy density filling up the whole spacetime. 
\pagebreak
The energy momentum tensor is just a constant diagonal one:
\eq
T_{\mu\nu}\! = \!
\left(\begin{tabular}{cccc}
$E$ & 0 &0 & 0 \\
0 & $p$ & 0 & 0 \\
0 & 0 & $p$ & 0 \\
0 & 0 & 0 & $p$
\end{tabular}\right)
\eqx 
with $E=3p$. In order to find the dual gravity background, we have to solve Einstein's equations with the boundary conditions given by 
(\ref{jane.bci})-(\ref{jane.bcii}). Due to the fact that the energy momentum tensor is constant, the metric will only depend on the $z$ coordinate and the Einstein's equations reduce to ordinary differential equations which can be solved explicitly. The result is
\eq
\label{jane.bhfg}
ds^2=-\f{(1-z^4/z_0^4)^2}{(1+z^4/z_0^4)z^2}\ dt^2
+(1+z^4/z_0^4)\f{dx^2_i}{z^2}+ \f{dz^2}{z^2}
\eqx
where the parameter $z_0$ is related to $E$ through
\eq
E=\f{3 N_c^2}{2\pi^2 z_0^4}
\eqx
Although it is not evident at first glance, the geometry (\ref{jane.bhfg}) is exactly the standard \emph{AdS planar black hole} \cite{wittenthermal}, but written in the Fefferman-Graham system of coordinates. We will give the explicit form of the coordinate transformation to the standard AdS Schwarzschild form shortly.
\index{AdS planar black hole}

The fact that the dual geometry turns out to be a black hole has significant implications for the physics. Let us note that we did not \emph{assume} that a black hole would appear. It came, in a unique way, from solving Einstein's equations with appropriate boundary conditions.

The parameter $z_0$ appearing in (\ref{jane.bhfg}) is the location of the black hole horizon. The Hawking temperature $T_H$ which is given by
\eq
T_H=\f{\sqrt{2}}{\pi z_0}
\eqx 
is identified with the gauge theory temperature. This may be most easily seen by computing the Hawking temperature through a Wick rotation of the metric (\ref{jane.bhfg}) and requiring the absence of a conical singularity at $z=z_0$. This requirement leads to a specific periodicity condition for the Euclidean time coordinate which is inversely proportional to the Hawking temperature. But, according to the AdS/CFT correspondence the metric induced on the boundary $z=0$ (up to an overall rescaling by~$z^2$) is exactly the metric of the (now Euclidean) gauge theory. Thus the gauge theory also has a compactified Euclidean time with the radius given by the same temperature 
\cite{wittenthermal}.

Another gravitational concept which carries over to the dual gauge theory is the Bekenstein-Hawking entropy which is identified with the entropy of the dual gauge theory plasma system. In this case, the entropy per spatial 3-volume is
\eq
s=\f{1}{4G_N} Area = \f{N_c^2}{2\pi} \left(\f{\sqrt{2}}{z_0}\right)^3
\eqx
Now we can use the relation between the horizon parameter $z_0$ and temperature to express the result completely in terms of gauge theoretical quantities
\eq
s=\f{1}{2} N_c^2 \pi^2 T^3
\eqx

Finally, we may use the relation between the energy density $E$ and $z_0$, and the link with temperature to express the energy density as a function of $T$. We get
\eq
E=\f{3}{8} N_c^2 \pi^2 T^4
\eqx
The nontrivial factor here is the numerical coefficient, which is \emph{different} from the corresponding one for the free massless gas (Stefan-Boltzmann). Similarly, the entropy density derived earlier is $3/4$ of the free gas answer \cite{janentropy34}. This mismatch is quite natural since here we are dealing with a strongly coupled plasma. In fact similar deviations from the Stefan-Boltzmann answer have been observed in lattice studies of QCD thermodynamics above the confinement/deconfinement phase transition.

Before we move on to discuss various systems of coordinates for this geometry,  let us note that it is exactly this geometry which is used to describe $\nn=4$ SYM at fixed nonzero temperature $T$. This interpretation is obvious from the above mentioned Euliclidean continuation, but can also be understood directly in Minkowski signature, where a link with the real-time formalism of finite-temperature QFT appears 
\cite{janmaldacenbh,janshenkerbh,jansonmink,janskenderismink}.

\subsubsection*{Various coordinate systems}

The geometry (\ref{jane.bhfg}) has been presented in the Fefferman-Graham coordinates, in which the connection to the gauge theory energy-momentum tensor is simplest. However these coordinates have also some significant drawbacks, of which one has to be aware.

Let us first perform a coordinate transformation to bring the metric 
(\ref{jane.bhfg}) into the standard AdS Schwarzschild form.
To this end set
\eq
\label{jane.fgstd}
z_{std}= \f{z}{\sqrt{1+{z^4}/{z_0^4}}}
\eqx
\pagebreak

\noindent{}Then, the metric becomes
\eq
ds^2= -\f{1-{z_{std}^4}/{\tilde{z}_0^4}}{z_{std}^2}\ dt^2
+\f{dx^2_i}{z_{std}^2}+ \f{1}{1-{z_{std}^4}/{\tilde{z}_0^4}}\f{dz^2}{z_{std}^2}
\eqx
with $\tilde{z}_0=z_0/\sqrt{2}$. Looking at the transformation of coordinates (\ref{jane.fgstd}), we see that the Fefferman-Graham coordinates cover only the region between the boundary and the horizon. Even if one would analytically continue the metric for $z>z_0$ one does not go beyond the horizon but rather returns back to the boundary.

The standard Schwarzschild coordinates also break down at the horizon and, in order to have explicit regularity at the horizon, it is convenient to introduce yet another system of coordinates --- the (ingoing) Eddington-Finkelstein coordinates.

These coordinates may be obtained from the standard Schwarzschild ones by redefining the time coordinate:
\eq
\label{jane.stdef}
t_{EF}=t-\f{1}{4} \tilde{z}_0 \left( 2 \arctan \f{z_{std}}{\tilde{z}_0} +
\log \f{\tilde{z}_0+z_{std}}{\tilde{z}_0-z_{std}} \right)
\eqx
The metric becomes then
\eq
ds^2=-\f{1-{z_{std}^4}/{\tilde{z}_0^4}}{z_{std}^2}\ dt^2_{EF}
+2 \f{dt_{EF} dz_{std}}{z_{std}^2}
+\f{dx^2_i}{z_{std}^2} 
\eqx
The crucial advantage of these coordinates is that the horizon is a perfectly regular point and one can enter the region inside the horizon. The lines 
$x^\mu=const^\mu$ are null geodesics falling into the black hole and reaching the singularity at $z_{std}=\infty$. These coordinates were used extensively in V. Hubeny's lectures at this school with $z_{std}$ substituted by 
$r=1/z_{std}$, which brings them to the canonical form.
\index{Eddington-Finkelstein coordinates}

Finally let us note that in the formula (\ref{jane.stdef}), the time coordinate gets an infinite shift at the horizon. In the case of the static black hole geometry this is completely harmless as the metric is time-independent, however for the time dependent geometries which will be the focus of these lectures this shift will give rise to some spurious singularities in the Fefferman-Graham treatment (fortunately appearing only at NNNLO in the large proper time expansion of the geometry).

Some comments are in order here. Of course in General Relativity nothing depends on the choice of coordinate system. This is true if we are dealing with an exact solution of Einstein's equations -- we may analyze it in any coordinate system we like. However if we perform an expansion in time in some coordinate system and deal with approximate solutions truncated at some order, we may get spurious singularities like in Fefferman-Graham at third order.

In these lectures we will nevertheless present the analysis in Fefferman-Graham coordinates (apart from a brief review of the Eddington-Finkelstein approach of \cite{janminwalla1} in section \ref{jans9}). The general formulation in Eddington-Finkelstein is considered in detail in the lectures of V. Hubeny, and has been applied to the boost invariant setting in \cite{janefboost1} and 
\cite{janefboost2,janefboost3}. One motivation for this choice of presentation is that our main focus is in reaching the small proper time regime, where we deal with 
\emph{exact} solutions of the Einstein's equations and hence do not need to worry about these subtleties. Also there, the analysis of initial conditions in Fefferman-Graham coordinates is simpler. 

\subsection{A case study: a planar shock wave}
\index{Shock wave, planar}

Another case of a gauge theory energy-momentum tensor for which the dual geometry is exactly solvable is a planar shockwave concentrated on the boundary.
The only nonvanishing component of $T_{\mu\nu}$ is 
\eq
T_{--}=\mu f(x^-)
\eqx
Such a configuration, for $f(x^-)=\mu \dl(x^-)$ represents a planar shock wave of gauge theory matter moving at the speed of light along one light cone direction. It may be understood to represent an analog of an ultrarelativistic nuclei.
Then the dual metric is found to be
\eq
ds^2=\f{dx^+ dx^- + f(x^-) z^4 {dx^-}^2 +dx_\perp^2+dz^2}{z^2}
\eqx
This configuration was proposed in \cite{janUS1}, being the simplest member of a family of shock wave solutions with $x_\perp$ dependence derived in 
\cite{janswgen} (see also \cite{janbeuf}).
A natural question to consider is a collision of two such shock waves, one propagating along the $x^-$ light cone direction, the other along $x^+$. There have been some preliminary investigation along these lines in 
\cite{janGR,jankovsw1,janshuryaksw}. However, a complete analysis remains to be done. Finally, let us note that this kind of shock wave is \emph{sourceless} in the bulk, in contrast to the shock waves considered recently in \cite{jangubsersw} and in the lectures of A. Yarom \cite{janYaromlect}. For references on work done on these other kinds of shock waves consult \cite{janYaromlect}.

\section{Boost-invariant flow}
\label{jans6}
\index{Bjorken boost invariant flow}

Let us now concentrate on a concrete evolving plasma system. Since eventually we would like to solve exactly Einstein's equations, one has to introduce as much symmetries as possible to reduce the complexity of the task, at the same time allowing for nontrivial physics to intervene. A natural choice in the context of heavy ion collisions is the requirement of longitudinal boost invariance. This assumption was introduced by Bjorken \cite{janbjorken} back in 1983 to model ultrarelativistic collisions. Basically, the motivation is that at infinite energy, a finite boost along the collision axis would not modify the physics. This is certainly not an ideal approximation, however it is used in basically all hydrodynamic codes for modelling relativistic heavy-ion collisions at RHIC \cite{janhydro}. We will make here an additional assumption that there is no dependence on the transverse coordinates, which corresponds to the limit of infinitely large nuclei. This is not really necessary for discussing the hydrodynamic limit (see \cite{janminwalla1,janhubenylect}) but will be essential in the far from equilibrium regime of small proper times.

When assuming boost invariance, it is natural to pass to proper-time/spacetime rapidity coordinates $(\tau,y,x_1.x_2)$
\eq
t=\tau \cosh y \qqqq x_3 = \tau \sinh y
\eqx
Then it turns out that the only non-vanishing components of the energy-momentum tensor are 
$T_{\tau\tau}$, $T_{yy}$ and $T_{xx} \equiv T_{x_1x_1}= T_{x_2x_2}$. Moreover, these components become functions of $\tau$ alone.

We should now impose tracelessness $T^\mu_\mu=0$ and conservation of energy momentum $T^{\mu\nu}_{;\nu}=0$ condtions, which take the form
\begin{align*}
-T_{\tau\tau}+\f{1}{\tau^2} T_{yy}+2T_{xx}&=0 \\
\tau \f{d}{d\tau} T_{\tau\tau} +T_{\tau\tau}+\f{1}{\tau^2} T_{yy}&=0
\end{align*}
These equations determine $T_{\mu\nu}$ uniquely in terms of a single function $\eps(\tau)$
\eq
\label{jane.tmunugenboost}
T_{\mu\nu}\! = \!
\left(\begin{tabular}{cccc}
$\eps(\tau)$ & 0 &0 & 0 \\
0 & $-\tau^3 \f{d}{d\tau} \eps(\tau)\!-\!\tau^2 \eps(\tau)$ & 0 & 0 \\
0 & 0 & $\eps(\tau)\!+\! \f{1}{2}\tau \f{d}{d\tau} \eps(\tau)$ & 0 \\
0 & 0 & 0 & $\eps(\tau)\!+\! \f{1}{2}\tau \f{d}{d\tau} \eps(\tau)$
\end{tabular}\right)
\eqx
The remaining function $\eps(\tau)$ can be interpreted as the energy density of the plasma at mid-rapidity (i.e. at $x_3=0$) as a function of (proper-) time.

Let us note that the above decomposition was purely `kinematical' --- valid in \emph{any} conformal 4D theory at any coupling. The determination of $\eps(\tau)$ will be an issue of understanding the dynamics of the theory of interest --- here $\nn=4$ SYM. In particular, weak coupling perturbative considerations \cite{jankovcgc} lead to the free streaming behaviour
\eq
\eps(\tau) \sim \f{1}{\tau}
\eqx
If, on the other hand, we would suppose that the plasma system behaves as a perfect fluid, then on top of the decomposition (\ref{jane.tmunugenboost}) we would impose
\eq
T^{\mu\nu}= (\eps+p)u^{\mu}u^{\nu} - p \eta^{\mu\nu}
\eqx
with $\eps=3p$. By our symmetry assumptions $u^\mu=(1,0,0,0)$ and we get in particular $p=\f{1}{\tau^2} T_{yy}=T_{xx}$, which gives a differential equation for $\eps(\tau)$
\eq
-\tau \f{d}{d\tau} \eps(\tau)\!-\!\eps(\tau) = 
\eps(\tau)\!+\! \f{1}{2}\tau \f{d}{d\tau} \eps(\tau)
\eqx

\pagebreak
\noindent{}with the celebrated Bjorken solution
\eq
\label{jane.epsperfect}
\eps(\tau)=\f{const.}{\tau^{\red{\f{4}{3}}}}
\eqx
Other dynamical assumptions would modify the functional form of $\eps(\tau)$. E.g. if the fluid would not be a perfect fluid but would have a nonzero viscosity (proportional to $T^3 \sim \eps^{\f{3}{4}}$ as should be the case for a conformal theory), then 
(\ref{jane.epsperfect}) would no longer be exact but would have corrections starting with
\eq
\eps(\tau)=\f{1}{\tau^{\f{4}{3}}} \left( 1-\f{2
  \eta_0}{\tau^{ \f{2}{3}}} +\ldots \right)
\eqx
with $\eta_0$ related to the shear viscosity. Here we set a single dimensional scale to unity. It can be easily reinstated in all terms by dimensional analysis.
Further $1/\tau^{4/3}$ corrections are  uniquely determined in terms of 
$\eta_0$. If the dynamics would follow $2^{nd}$ order viscous hydrodynamics, these $1/\tau^{4/3}$ corrections would be \emph{different} and would involve additional, second order transport coefficients.

So we see that the knowledge of $\eps(\tau)$ contains a lot of information on the dynamics of plasma. In the rest of these lectures our goal will be to deduce what $\eps(\tau)$ is singled out by the AdS/CFT correspondence. Initially we will concentrate on the large $\tau$ asymptotics of $\eps(\tau)$ and then, in section \ref{jans11}, move to consider the behaviour of $\eps(\tau)$ for small $\tau$.

\section{Large proper time behaviour}
\label{jans7}

Let us first concentrate on the large $\tau$ asymptotics of $\eps(\tau)$ and consider determining the exponent $s$ in
\eq
\label{jane.epstaus}
\eps(\tau) \sim \f{1}{\tau^s} \qqqq \text{for $\tau\to \infty$}
\eqx 
We will follow here the strategy outlined in section \ref{jans4}, and construct, for each $s$, the dual geometry. Then we will check for which $s$ the dual geometry is nonsingular. This condition will determine $s$. This approach was proposed in \cite{janUS1}, where more details may be found.
But first, let us narrow down the range of possible $s$. We will demand that 
the energy density in any reference frame is non-negative i.e.
\eq
T_{\mu\nu}t^\mu t^\nu \geq 0
\eqx
for any timelike 4-vector $t^\mu$. This leads to the inequalities
\eq
\eps(\tau)\geq 0 \qqqq \eps'(\tau)\leq 0 \qqqq
\tau \eps'(\tau) \geq -4 \eps(\tau)
\eqx
In particular, for (\ref{jane.epstaus}), we obtain that $0 \leq s \leq 4$.

\subsection{The AdS/CFT analysis}

Now we have to construct the dual geometry to a plasma configuration with the energy density behaving like (\ref{jane.epstaus}). Since the geometry will have the same symmetries as assumed for the plasma system, we are led to the following ansatz
\eq
\label{jane.boostansatz}
ds^2=\f{1}{z^2} \left( -e^{a(z,\tau)} d\tau^2+ e^{b(z,\tau)} \tau^2 dy^2+ e^{c(z,\tau)} dx^2_\perp \right) +\f{dz^2}{z^2}
\eqx
Again, let us reiterate that the above ansatz is completely general. At this stage the choice of the Fefferman-Graham system of coordinates is perfectly legitimate. 

According to the approach explained in section \ref{jans4}, we have to solve Einstein's equations 
\eq
R_{\al\bt}-\f{1}{2}g^{5D}_{\al\bt} R - 6\, g^{5D}_{\al\bt}=0
\eqx
with the boundary conditions 
\eq
\label{jane.boostbc}
a(z,\tau)=-z^4 \eps(\tau) +z^6 a_6(\tau)+ z^8 a_8(\tau) +\ldots
\eqx
It is instructive to find the explicit form of the first few coefficients in the above Taylor series\footnote{This is an \emph{exact} result without any approximation.}. Using a computer algebra system we obtain
\eqn
a\left( \tau, z \right)\!\!\! &=&\!\!\! - \eps (\tau ) \cdot z^4 + \left\{-\frac{\eps '(\tau
   )}{4 \tau }-\frac{\eps ''(\tau )}{12}\right\} \cdot z^6 + \left\{ \frac{1}{6} \eps (\tau )^2+\frac{1}{6} \tau  \eps '(\tau )
   \eps (\tau )+\right.\nonumber\\
   &&\!\!\!\!\!\!\!\!\!\!\!\!\!\!\! \left.
+\frac{1}{16} \tau ^2 \eps '(\tau)^2+\frac{\eps '(\tau )}{128 \tau ^3}-\frac{\eps ''(\tau
   )}{128 \tau ^2}-\frac{\eps ^{(3)}(\tau )}{64 \tau }-\frac{1}{384}
   \eps ^{(4)}(\tau ) \right\} \cdot z^{8} + \cdots
\eqnx
Let us now specialize to the case of interest $\eps(\tau)=1/\tau^s$. We get
\begin{align*}
-&{z}^{4}\ \red{{\tau}^{-s}} + {z}^{6}\ \left( \f{1}{6}\,{\tau}^{-s-2}s-
\f{1}{12}\,
{\tau} ^{-s-2}{s}^{2} \right)+\\
+&{z}^{8}\ \left( {\bf
-\f{1}{16}\,{\tau}^{-2\,s}{s}^{2}-\f{1}{6}\,{\tau}^{-2\,s}+\f{1}{6}\,
{\tau}^{-2\,s}s}+{\frac {1}{96}}\,{\tau}^{-s-4}{s}^{2}-{\frac {1}{384}}\,{\tau}
^{-s-4}{s}^{4} \right) + \ldots
\end{align*}
where the terms dominant for large $\tau$ are outlined in bold.
Looking at the above formula and analyzing a couple of higher order terms we may convince ourselves that the dominant terms in 
$a_{n}(\tau)$ for large $\tau$ will be of the form 
\eq
z^{n} a_{n}(\tau) \sim  \f{z^n}{\tau^{\f{n s}{4}}} = \left( 
\f{z}{\tau^{\f{s}{4}}} \right)^n  \qqqq \text{ for large $\tau$}
\eqx
This shows that it is natural to introduce a scaling variable
\eq
v \equiv \f{z}{\tau^{\f{s}{4}}}
\eqx
Consequently, the metric coefficients will have an expansion of the form
\eq
a(z,\tau) =a_0(v) +\f{1}{\tau^{\#}} a_1(v)+ \ldots
\eqx
Several comments are in order here. Firstly, the appearance of the scaling variable at late times is a dynamical consequence of the structure of Einstein's equations. We will find later, that for small proper times a similar structure will \emph{not} appear\footnote{For the case $\eps(\tau)\to const$ as $\tau\to 0$.}. Secondly, the separation of dynamics into a scaling variable and an expansion in inverse powers of $\tau$ corresponds to a gradient expansion (c.f. \cite{janminwalla1} and the lecture by V. Hubeny \cite{janhubenylect}). Finally, the appearance of a scaling variable reduces the very complicated Einstein's equations to a system of nonlinear \emph{ordinary} differential equations:
\begin{align*}
 v(2 a'(v)c'(v)+a'(v)b'(v)+2 b'(v) c'(v)) -6a'(v)-6b'(v)
-12 c'(v)+ v c'(v)^2 &=0 \\
 3v c'(v)^2+v b'(v)^2+2v b''(v)+4v c''(v)-6b'(v)-12 c'(v)+2v b'(v)
c'(v) &=0 \\
 2 v \red{s} b''(v) +2 \red{s} b'(v)+8 a'(v)-v \red{s} a'(v) b'(v) -8b'(v) +v \red{s} b'(v)^2 + \hspace{1cm} &\\
 +4 v \red{s} c''(v) +4 \red{s} c'(v) -2 v \red{s} a'(v) c'(v) +2 v \red{s} c'(v)^2 &=0 
\end{align*}
which can be solved exactly. The solution is
\begin{align*}
a(v) &= A(v)-2m(v)  \\
b(v) &= A(v)+(2\red{s}-2) m(v) \\
c(v) &= A(v)+(2-\red{s}) m(v)
\end{align*}
where
\begin{align*}
A(v)=&\f{1}{2} \left( \log(1+\red{\Dl(s)}\,v^4) +\log(1-\red{\Dl(s)}\, v^4) \right) \\
m(v)=&\f{1}{4\Dl(s)} \left( \log(1+\red{\Dl(s)}\,v^4) -\log(1-\red{\Dl(s)}\, v^4)
\right)
\end{align*}
with
\[
\Dl(s)=\sqrt{\f{3s^2-8s+8}{24}}
\]
Now we can analyze the singularities of the above geometry. We see that there is a potential singularity where the argument of the logarithm vanishes. Of course, it may well be a coordinate singularity, so we have to evaluate a curvature invariant like
\eq
\rsq=R^{\mu\nu\alpha\beta}R_{\mu\nu\alpha\beta}
\eqx
Moreover, since our geometry is only exact in the scaling limit, we have to evaluate $\rsq$ in the same limit i.e. $\tau \to \infty$, $z\to \infty$, keeping the ratio $v=\f{z}{\tau^{\f{s}{4}}}$ fixed. 
The resulting expression is quite complicated and can be found in \cite{janUS1}. Its general structure is
\eq
\rsq=\frac{Numerator(v;s)}{ \left( 1-\Dl(s)   ^{2}{v}^{8} \right) ^{4}}
\eqx
We see that for generic $s$, there is a $4^{th}$ order pole singularity in the curvature. It turns out that this singularity gets cancelled by the numerator \emph{only} for a single value of $s$:
\eq
s=\f{4}{3}
\eqx
which is just the asymptotic scaling characteristic of perfect fluid hydrodynamics. In this way we see that nonlinear perfect fluid hydrodynamics arises at late stages of plasma expansion as a consequence of the AdS/CFT correspondence.

\begin{figure}
\label{janfig.fgef}
\centerline{\includegraphics[height=4cm]{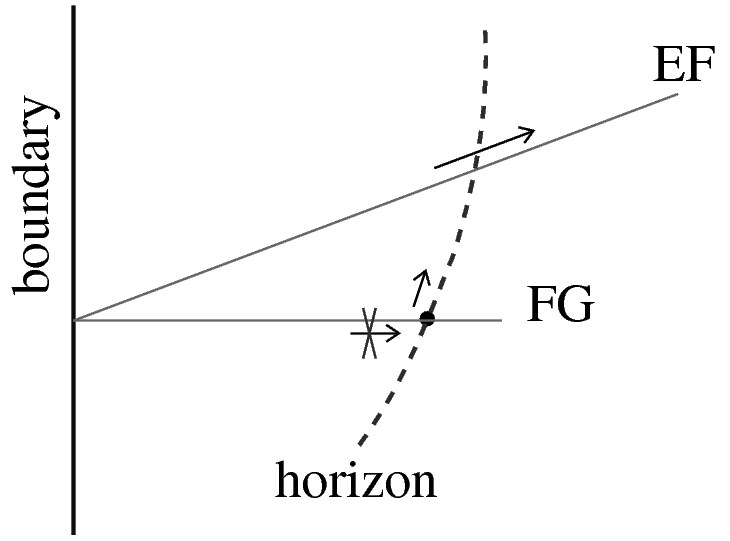}}
\caption{The difference between using Fefferman-Graham and Eddington-Finkelstein coordinates for checking nonsingularity represented by arrows.}
\end{figure}

Note that in this way we do not approach the `horizon' directly, but approach it asymptotically along a `parallel' trajectory. An analogous analysis using Eddington-Finkelstein coordinates enables us to pass through the `horizon' and require directly the nonsingularity of the metric there. The difference between the two procedures is summarized in 
Fig.~1. Both methods give equivalent results up to two subleading 
orders in the large proper-time expansion. In order to go beyond that, however, one has to use Eddington-Finkelstein coordinates (currently the only result in this direction beyond $2^{nd}$ order is in \cite{janboothmh} in the boost-invariant setting).

\subsection{Perfect fluid geometry}

Let us now examine more closely the dual geometry corresponding to the perfect fluid value of $s=4/3$. Then the complicated formulas for the metric coefficients involving generically irrational powers and square roots obtained above simplify drastically and we obtain\footnote{We reinstated here a trivial dimensional scale $e_0$.}
\eq
ds^2=\f{1}{z^2} \left[- \f{\left( 1-\green{\f{e_0}{3}}
      \f{z^4}{\tau^{4/3}} \right)^2}{1+\green{\f{e_0}{3}}
\f{z^4}{ \green{\tau^{4/3}}}} d\tau^2+
\left(  1+\green{\f{e_0}{3}} \f{z^4}{\green{\tau^{4/3}}}\right) (\tau^2 dy^2 +dx^2_\perp)\right] 
+ \f{dz^2} {z^2}
\eqx
This geometry is analogous to a black hole (c.f. (\ref{jane.bhfg})) with the position of the horizon moving into the bulk as
\eq
z_0=\sqrt[4]{\f{3}{e_0}} \cdot \tau^{\f{1}{3}}
\eqx
This has a clear physical interpretation. Recall that for a static black hole, the position of the horizon in the bulk is inversely proportional to the temperature. Thus here we have a dual counterpart of the plasma undergoing cooling during expansion\footnote{The dual counterpart of cooling was first suggested qualitatively in \cite{janSZSJS}.}. Indeed naively generalizing the static formulas leads to 
\eq
T = \f{\sqrt{2}}{\pi z_0}= \f{2^{\f{1}{2}} e_0^{\f{1}{4}}}{\pi 3^{\f{1}{4}}} 
\tau^{-\f{1}{3}}
\eqx
A more indepth analysis of these phenomenae using the framework of event 
\cite{janmukundevent} or dynamical \cite{janefboost2,janmukunddyn,janboothmh} horizons has been made, although a complete understanding of the notions of temperature and entropy in the fully dynamical case is still lacking.

\section{Plasma dynamics beyond perfect fluid}
\label{jans8}

One of the key discoveries of the AdS/CFT correspondence was the derivation of the universal value of shear viscosity for plasma at finite nonzero temperature. This was done using linear response theory in \cite{janson}. It is thus interesting to examine how do viscous effects manifest themselves in the current nonlinear setting.

Let us first examine the question whether we can see if the perfect fluid dynamics is violated from the dual gravitational point of view. Suppose that it is not and that consequently
\eq
\label{jane.epspfluid}
\eps(\tau)=\f{1}{\tau^{\f{4}{3}}}
\eqx
is valid at all proper times without any corrections. Then one can find the next orders in the large $\tau$ scaling expansion of the metric coefficients
\eq
a(z,\tau) =a_0(v)+\f{1}{\tau^{\red{\f{4}{3}}}} a_2(v)+ \ldots 
\eqx
This can be done explicitly since, fortunately, the equations for the corrections are linear albeit still quite complicated.
Peforming this calculation, and computing the curvature $\rsq$ up to this order yields
\eq
\rsq=R_{\al\bt\gm\dl}R^{\al\bt\gm\dl}=\underbrace{R_0(v)}_{nonsingular}+
\f{1}{\tau^{\red{\f{4}{3}}}} \underbrace{R_2(v)}_{\red{singular!}}+\ldots
\eqx
where the indicated singularity is of the very strong $4^{th}$ order pole type. This means that the perfect fluid behaviour (\ref{jane.epspfluid}) has to be violated.

Let us now be completely agnostic about viscous hydrodynamics and assume a completely generic type of corrections:
\eq
\label{jane.epsgencorr}
\eps(\tau)=\f{1}{\tau^{\red{\f{4}{3}}}} \left(
1-\f{2\green{A}}{\tau^{\green{r}}} \right) 
\eqx
Solving the Einstein's equations with the appropriate boundary conditions set by~(\ref{jane.epsgencorr}), computing the curvature\footnote{Various steps of this calculation were done in \cite{janshinvisc}, \cite{janrjvisc} and \cite{janhellermsc}.} yields
\eq
\rsq=R_{\al\bt\gm\dl}R^{\al\bt\gm\dl}=\underbrace{R_0(v)}_{nonsingular}+
\f{1}{\tau^{\green{r}}} \underbrace{R_1(v)}_{nonsingular}+
\f{1}{\tau^{\green{2r}}} \underbrace{\tilde{R}_2(v)}_{\red{singular!}}+
\f{1}{\tau^{\red{\f{4}{3}}}} \underbrace{R_2(v)}_{\red{singular!}}+\ldots
\eqx
The last two terms are always singular. The only way that we may obtain bounded curvature is to make those two terms cancel between themselves. This requires 
\eq
r=\f{2}{3}
\eqx
which is exactly the correct scaling for a correction coming from shear viscosity. Moreover, we have to fine tune the coefficient $A$ to \cite{janrjvisc}
\eq
A=2^{-\f{1}{2}} 3^{-\f{3}{4}}
\eqx
which is the value corresponding to the value of the shear viscosity to entropy ratio
\eq
\f{\eta}{s}= \f{1}{4\pi}
\eqx
In this way we reconfirmed, in a fully nonlinear setting \cite{janrjvisc}, the value of shear viscosity computed at fixed temperature \cite{janson}.

The above analysis can be repeated for the NNLO correction \cite{janrjmhsecond} with the final result for $\eps(\tau)$:
\eq
\label{jane.hydrexp}
\eps(\tau)=\f{1}{\tau^{\red{\f{4}{3}}}} -\green{\f{2}{2^{\f{1}{2}} 3^{\f{3}{4}}}} \f{1}{\tau^{\red{2}}} +
 \green{\f{1+2\log 2}{12 \sqrt{3}}} \f{1}{\tau^{\red{\f{8}{3}}}} +\ldots
\eqx
The coefficient of the NNLO term involves $2^{nd}$ order transport coefficients.
It is at this stage that the pathologies of Fefferman-Graham coordinates surface, leading to a leftover logarithmic singularity in the scaling limit of the curvature (appearing at NNNLO in the metric, which is necessary for obtaining the coefficients of $\eps(\tau)$ at one order lower). The singularity was found to be persistent and not associated with truncating other fields of ten dimensional supergravity \cite{janusbuchel}.  
Its origin has been explained in detail in \cite{janefboost1} (see also \cite{janefboost2,janefboost3}) and can be associated with the singular transformation between coordinates regular at the horizon (Eddington-Finkelstein) and the Fefferman-Graham ones, coupled with performing an expansion of the geometry w.r.t. those coordinates.
In order to proceed further, which is however rather impractical analytically, one would have to perform the analysis in Eddington-Finkelstein coordinates (a result in this direction, the NNNLO term in $\eps(\tau)$ is given in \cite{janboothmh}).

\section{Interlude: hydrodynamics redux}
\label{jans9}

In the above, we have adopted a very agnostic attitude towards the expected dynamics ruling the time evolution of the energy-momentum tensor of the $\nn=4$ plasma system. We did not assume that one could parametrize the $T_{\mu\nu}$ in terms of such quantities as flow velocity and energy/pressure. We started off from the most general $T_{\mu\nu}$ consistent with the assumed symmetries. By proceeding in this way we have an option of describing dynamics which does not fit at all into a hydrodynamical language. This is in fact the main reason for presenting such an approach here, as in the remaining part of the lectures we would like to address the dynamics of boost-invariant plasma at small proper times where we do not expect hydrodynamic description to be a good starting point.

On the other hand such flexibility has also significant drawbacks. The determination of the transport coefficients of hydrodynamics presented above followed by first deriving from AdS/CFT the explicit form of $\eps(\tau)$. Given that $\eps(\tau)$, one could ascertain that the leading term is a solution of perfect fluid equations of motion, and together with the subleading term is a solution of viscous hydrodynamic equations with a specific value of the shear viscosity (we leave this as an exercise for the reader).   

It is thus very interesting to obtain directly the hydrodynamic equations from AdS/CFT without making the passage through explicit solutions. This task was performed in \cite{janminwalla1} and is presented in detail in the lectures of V. Hubeny at this school \cite{janhubenylect}. For completeness, let us just summarize here the main idea.

The static black hole geometry presented in section \ref{jans4} is dual to a plasma at rest, which can be described by a flow vector $u^\mu=(1,0,0,0)$ and an energy density $E$ (or equivalently temperature $T$). By performing a boost one can obtain a dual geometry to a uniformly moving plasma with 4-velocity $u^\mu$. 

\pagebreak 
\noindent{}In Eddington-Finkelstein coordinates it is given explicitly as
\eq
\label{jane.boostbh}
ds^2=-2u_\mu dx^\mu dr-r^2 \left(1- \f{T^4}{\pi^4 r^4} \right)u_\mu
u_\nu dx^\mu dx^\nu +r^2(\eta_{\mu \nu}+u_\mu u_\nu)  dx^\mu dx^\nu
\eqx
where $r=\infty$ corresponds to the boundary, $r=T/\pi$ is the horizon
while $r=0$ is the position of the singularity ($r=1/z_{EF}$ c.f. section \ref{jans4}).
This is an exact solution of Einstein's equations. Now promote $T$ and $u^\mu$ to slowly varying functions of the boundary Minkowski coordinates. 
The geometry (\ref{jane.boostbh}) ceases to be a solution of Einstein's equations and has to be corrected by terms proportional to gradients. These correction terms can be determined with the integration constants fixed by the requirement of nonsingularity at the horizon.

Now from the corrected geometry one can read off the $T_{\mu\nu}$ which is explicitly expressed (similarly to the metric) in terms of $T$, $u^\mu$ and the gradients of $u^\mu$. The numerical constants coming from nonsingularity become exactly the transport coefficients. In this way one obtains
\begin{align}
T^{\mu\nu}_{rescaled}&=\underbrace{(\pi T)^4 (\eta^{\mu\nu}+4u^\mu
  u^\nu)}_{\black{perfect\ fluid}}- \underbrace{2(\pi T)^3
  \sg^{\mu\nu}}_{\black{viscosity}}+ \nonumber\\ 
&+ \underbrace{(\pi T^2)\left( \log 2 T^{\mu\nu}_{2a}+2 T^{\mu\nu}_{2b}+(2-\log
2) \left( \f{1}{3} T^{\mu\nu}_{2c}+T^{\mu\nu}_{2d}+T^{\mu\nu}_{2e}
\right)\right)}_{\black{second\ order\ hydrodynamics}} \nonumber
\end{align}
The energy-momentum conservation $\partial_\mu T^{\mu\nu}=0$ of such a 
$T^{\mu\nu}$  is by definition the hydrodynamic relativistic Navier-Stokes equation.

From the above construction we see that the appearance of hydrodynamics in the AdS/CFT correspondence is now completely understood. For any solution of the hydrodynamic equations $T(x^\rho)$, $u^\mu(x^\rho)$, the formula 
(\ref{jane.boostbh}) and its correction terms give an explicit dual metric valid to the same order of the derivative expansion. A similar analysis was performed later in the Fefferman-Graham coordinates \cite{janfggen}.

One final thing to note is that the starting point in the above construction is the boosted black hole, which means that one assumes that approximately one is dealing with an energy-momentum tensor of a hydrodynamic type (i.e. parametrizable by a flow velocity and energy density). This does not always need to be the case, as we shall see shortly, and then one has to return to an \emph{ab-initio} analysis of Einstein's equations of the type presented in section \ref{jans7}.

\section{Plasma dynamics beyond hydrodynamics}
\label{jans10}

The appearance of hydrodynamic behaviour of strongly coupled plasma as a consequence of the AdS/CFT correspondence is certainly very interesting and satisfying theoretically, however perhaps the most fascinating feature of AdS/CFT is its ability to address the behaviour of a plasma system very far from equilibrium, where in QCD we do not even have a well motivated phenomenological model.

As an example of a configuration which cannot be described, even in any approximation, by hydrodynamic methods consider the problem of plasma isotropisation, extensively studied in various variations within QCD 
\cite{janMrowczynski,janYaffe,janRomStrik}. 

Suppose we have a plasma system uniform in space  with anisotropic pressures. In weakly coupled gauge theory one could consider a gas of gluons with non isotropic momentum distributions, like gaussians with different widths for the different momentum components. Then one expects that the pressures would isotropise in time. The energy-momentum tensor of such a system would have the form
\eq
\label{jane.iso}
T_{\mu\nu}=
\begin{pmatrix}
		\eps & 0 & 0 & 0 \\
		0 & p_\parallel(t) & 0 & 0 \\
		0 & 0 & p_\perp(t) & 0 \\
		0 & 0 & 0 & p_\perp(t) \\
\end{pmatrix}
\eqx
Note that such a system cannot be described by any form of (even all-order resummed) viscous hydrodynamics, since by symmetry $u^\mu=(1,0,0,0)$ and thus has vanishing derivatives. So all viscous terms vanish, while the leading term is clearly of a different form.
However nothing stops us from applying the AdS/CFT analysis using Einstein's equations to such a system. This has been first proposed in \cite{jananisous}. Subsequently, a numerical study of this system was performed in \cite{jananisocy}.

Another interesting problem, which we will discuss in the remaining part of these lectures, is the behaviour of the boost-invariant plasma system considered before but now at small proper times. Since the hydrodynamic expansion 
(\ref{jane.hydrexp})
\eq
\label{jane.hydrexpb}
\eps(\tau)=\f{1}{\tau^{\red{\f{4}{3}}}} -\green{\f{2}{2^{\f{1}{2}} 3^{\f{3}{4}}}} \f{1}{\tau^{\red{2}}} +
 \green{\f{1+2\log 2}{12 \sqrt{3}}} \f{1}{\tau^{\red{\f{8}{3}}}} +\ldots
\eqx
is an expansion in inverse powers of $\tau$, it completely breaks down as we approach $\tau=0$. Here the situation is more complicated than in the case of uniform isotropisation (\ref{jane.iso}) mentioned above as we expect a mixture of non-equilibrium and hydrodynamic behaviour. In fact it is exactly the question of the transition to hydrodynamics, and what factors are important in setting the scale of this transition, that is very interesting in the context of heavy-ion collisions. A related more general issue is the observation of thermalization (proposed in \cite{jannastase} to be related to a formation of a black hole in the bulk) and an analysis of the concrete way in which this scenario is realised.

\section{Dynamics at small proper time}
\label{jans11}

For the reasons described above, we will have to deal with the full Einstein's equations. From the point of view of hydrodynamics treated as a gradient expansion these encompass \emph{all orders} of viscous hydrodynamics together with an infinite set of higher transport coefficients. But apart from these there is additional information contained in the Einstein's equations. Taking the case of a planar black hole as an example, all order hydrodynamics may be identified, on the linearized level, with the lowest quasinormal mode and its exact dependence on spatial momentum. But apart from this lowest mode there is an infinite set of higher quasinormal modes which decay exponentially (in  the AdS/CFT context, see in particular  \cite{janqnmhor}). And all of these become important in a far from equilibrium situation such as the early time dynamics of the boost-invariant flow.

Here we will again go to the boost invariant setting and repeat the analysis
of section \ref{jans7}, but now concentrating on the small $\tau$ regime. 
We follow the analysis of \cite{janusearly}.
We will adopt the same ansatz for the metric (\ref{jane.boostansatz}) and solve Einstein's equations with the boundary conditions (\ref{jane.boostbc}). 

Before going into the details of this construction, let us comment why we are using the Fefferman-Graham system of coordinates. In contrast to the late time expansion, here we will aim at solving the Einstein's equations to an arbitrary accuracy --- without performing any kind of scaling limit. Therefore any choice of coordinates works equally well. Moreover the constraint equations for initial data in the Fefferman-Graham system of coordinates are particularly transparent.

Let us first determine the qualitative behaviour of $\eps(\tau)$ at small 
$\tau$. We will do it in two ways.

\subsection*{The absence of a scaling variable}

In the large proper time regime, the structure of Einstein's equations naturally led to the introduction of a scaling variable, which reduced the problem to solving ordinary differential equations and a subsequent expansion in inverse powers of $\tau$. Let us now analyze the solution of Einstein's equations at small $\tau$ from this point of view.

We again start from the \emph{exact} power series solution
\eqn
\label{jane.aexact}
a\left( \tau, z \right)\!\!\! &=&\!\!\! - \eps (\tau ) \cdot z^4 + \left\{-\frac{\eps '(\tau
   )}{4 \tau }-\frac{\eps ''(\tau )}{12}\right\} \cdot z^6 + \Big\{ \frac{1}{6} \eps (\tau )^2+\frac{1}{6} \tau  \eps '(\tau )
   \eps (\tau )+\nonumber\\
   &&\!\!\!\!\!\!\!\!\!\!\!\!\!\!\!
+\frac{1}{16} \tau ^2 \eps '(\tau)^2+\frac{\eps '(\tau )}{128 \tau ^3}-\frac{\eps ''(\tau
   )}{128 \tau ^2}-\frac{\eps ^{(3)}(\tau )}{64 \tau }-\frac{1}{384}
   \eps ^{(4)}(\tau ) \Big\} \cdot z^{8} + \cdots
\eqnx 
and substitute the asymptotics
\eq
\eps(\tau) \sim \f{1}{\tau^s} \qqqq \text{for $\tau\to 0$}
\eqx
In this way we obtain
\begin{align}
\label{jane.sersmall}
-&{z}^{4}\ {\tau}^{-s} + {z}^{6}\ \left( \f{1}{6}\,{\tau}^{-s-2}s-
\f{1}{12}\,
{\tau} ^{-s-2}{s}^{2} \right)+\nonumber\\
+&{z}^{8}\ \left( 
-\f{1}{16}\,{\tau}^{-2\,s}{s}^{2}-\f{1}{6}\,{\tau}^{-2\,s}+1/6\,
{\tau}^{-2\,s}s+\bf{{\frac {1}{96}}\,{\tau}^{-s-4}{s}^{2}-{\frac {1}{384}}\,{\tau}
^{-s-4}{s}^{4}} \right) + \ldots
\end{align}
where the terms dominating for small $\tau$ are rendered in bold. This analysis was first done by Kovchegov and Taliotis \cite{jankovearly}, who deduced that for {\emph generic} $s$ the dominant terms at small $\tau$ are of the form
\eq
\f{z^4}{\tau^s}\cdot f\left( \green{w\equiv\f{z}{\tau} }\right)
\eqx
In \cite{jankovearly}, the scaling solution was found, but due to its complex branch cut structure\footnote{And an additional physical bound on $s$, see 
\cite{jankovearly}.}, Kovchegov and Taliotis argued that the only acceptable solution had $s=0$, which is a very interesting result.

But if we again look at (\ref{jane.sersmall}), we see that the terms resummed by the scaling variable \emph{vanish} for $s=0$ and are no longer dominant. Hence there is no place for a scaling variable at small $\tau$ and for $s=0$ one has to reanalyze the Einstein equations in order to describe the full solution at 
$\tau \sim 0$.

\subsection*{The existence of a regular initial condition}

One can reach the same conclusion, as well as some more detailed information, on the small $\tau$ dependence of $\eps(\tau)$ \emph{assuming} that at $\tau=0$ we have a regular initial condition\footnote{This is an assumption which may, or may not be realistic for ultraenergetic collisions. We prefer to keep the options open and analyze boost invariant flow with regular initial conditions as an interesting nonequilibrium dynamical system for its own sake.}. Recall the expression (\ref{jane.aexact}) and subsitute $\tau=0$. Firstly, we see that the assumption that the metric coefficients are finite leads to a finite limit of $\eps(\tau)$ as $\tau \to 0$, consistent with $s=0$. Secondly, the inverse powers of $\tau$ appearing in the higher order terms do not lead to a singularity if and only if $\eps(\tau)$ has an expansion only in \emph{even} powers of $\tau$:
\eq
\eps(\tau)=\eps_0+\eps_2 \tau^2 +\eps_4 \tau^4+\ldots
\eqx

A closer analysis reveals that the coefficients $\eps_{2n}$ are uniquely determined, through the Einstein's equations, in terms of the coefficients of the initial condition for the metric:
\eq
\label{jane.ainitexp}
a(\tau=0,z)=a_0 z^4+ a_2 z^6+a_4 z^* +\ldots
\eqx
In order to complete the analysis of the gravitational setup, we have to analyze what are the admissible initial conditions (\ref{jane.ainitexp}). Once this is done, one can set up the analysis of the system by evolving the geometry from (\ref{jane.ainitexp}) using Einstein's equations and read off 
$\eps(\tau)$ from the metric. This can be done either numerically solving Einstein's equations \cite{janWIP}, or analytically by expressing 
the coefficients $\eps_{2n}$ directly in terms of the coefficients of the initial condition $a_{2n}$ \cite{janusearly}.

\subsection*{The classification of possible initial conditions}

As is well known, in General Relativity, the initial conditions cannot be arbitrary but have to satisfy some nonlinear constraint equations. This causes the gravitational initial value problem to be quite nontrivial in general.
Fortunately, for the case of the $\tau=0$ hypersurface in the Fefferman-Graham coordinates, the constraints can be solved exactly.

Let us denote by $E_{\al\bt}$, the components of Einstein's equations written in the form
\eq
E_{\al\bt} \equiv R_{\al\bt}+4g_{\al\bt}=0
\eqx
Then the constraints are contained in equations $E_{\tau z}$ and $E_{zz}$. Denoting $a_0(z) \equiv a(\tau=0,z)$ etc. we get at once
\eq
a_0(z)=b_0(z) \qqqq \dot{a}_0=\dot{b}_0=\dot{c}_0=0
\eqx
and the only remaining costraint is the single nonlinear equation
\eq
\label{jane.nonlini}
a_0''+ c_0''+\f{1}{2} (a'_0)^2 +\f{1}{2} (c'_0)^2- \f{1}{z} \left(a'_0 + c'_0\right)=0 
\eqx
Let us note an extremely surprising feature of the above equation. At the linearized level, it has a trivial solution $a_0(z)=-c_0(z)$. So one may expect that for infinitesimal $a_0$, the function $c_0$ would be also very small and only slightly deformed from $-a_0$ by taking into account the effect of the nonlinear terms.
It turns out, however, that this is never true, and the nonlinearity always causes a blowup of the solution for some finite $z$. To see this introduce $v(z^2) = \frac{1}{4z} a_0'(z)$ and similarly $w(z^2)$ for $c_0$. Then the constraint equation (\ref{jane.nonlini}) takes the simple form
\eq
\label{jane.nonlinii}
v'+ w'+v^2+ w^2=0
\eqx
Now it is easy to see, that there does not exist an everywhere bounded ($v=w=0$ at infinity) solutions of the constraint equations. To this end it is enough to integrate (\ref{jane.nonlinii}) to get
\eq
0=\int_0^\infty (v'+w') +\int_0^\infty (v^2+w^2)=\int_0^\infty (v^2+w^2)
\eqx
Therefore, at $\tau=0$ with boost invariant symmetry, gravity leads to inherently nonlinear dynamics --- a linearized regime does not exist at all!

It is not difficult to impose conditions on $v$ and $w$ so that the blowup is not a curvature singularity --- at least $\rsq$ stays finite there (see \cite{janusearly} for details). Moreover one can solve the constraint (\ref{jane.nonlinii}) analytically. Indeed, defining $v_+ = -w-v$, $v_- = w-v$,
\eq
v_-= \sqrt{2v_+'-v_+^2}
\eqx
solves (\ref{jane.nonlinii}) for any $v_+$.
Let us conclude with an example of a simple particular solution of the initial value constraints:
\eq
\label{jane.cosh}
a_0(z)=b_0(z)=2\log \cos a z^2 \qqqq c_0(z) = 2\log \cosh a z^2
\eqx

The huge range of initial conditions that can be imposed is in fact quite natural. On the gauge theory side we may also expect to be able to prepare an initial state with the same energy density in a multitude of ways. Let us contrast this with the large $\tau$ expansion (\ref{jane.hydrexpb}) of 
$\eps(\tau)$, which only depends on a \emph{single} scale\footnote{In (\ref{jane.hydrexpb}) this scale has been set to unity, but may be reinstated unambigously by dimensional analysis.}. In other words, once the dominant asymptotics of $\eps(\tau)$ is known, all subleading powerlike terms are uniquely determined.

The physical interpretation of this difference is quite clear. We expect dissipative effects to wash out differences in initial conditions leaving only a single scale (under the present symmetry assumptions) governing the large proper time expansion of $\eps(\tau)$. On the gravitational side, these effects can be understood as nonlinear generalizations of higher quasinormal modes which die off exponentially (see \cite{janboostqnm} for some analysis along these lines in the boost invariant setting).

\subsection*{An analysis of some aspects of the small proper time behaviour of~$\eps(\tau)$}

Once we have the allowed initial conditions at $\tau=0$, we need to solve Einstein's equations with these initial data. Then, as explained in section \ref{jans4}, we may read off $\eps(\tau)$ from the solution of Einstein's equations. Since we do not have a scaling variable at our disposal we have to do it exactly. The ideal way to proceed would be to solve Einstein's equations numerically. This study is currently under way \cite{janWIP}. The route followed in \cite{janusearly} was to solve these equations for the metric in a power series in $z$ and $\tau$, obtaining a power series expression for $\eps(\tau)$ up to the order 
$\tau^{100}$ for some initial conditions. A drawback of the above method is that the power series in question has a finite radius of convergence, necessitating the use of Pade approximations as an extrapolation method. 
Below we will present some analysis of these extrapolated profiles 
\cite{janusearly}. We have to emphasize, however, that one would require a real numerical solution of Einstein's equation to be sure of all the details.

First let us discuss the transition to hydrodynamics. One possibility of quantifying this is by considering an `effective exponent' of the power law dependence of $\eps(\tau)$ defined as
\eq
\label{jane.effpower}
-\tau \f{d}{d\tau} \log\, \eps(\tau)
\eqx
Initially it is zero, and it should rise up to $4/3$ for late time expansion. In order to evaluate it we plot a Pade approximant (of constant large $\tau$ asymptotics) of the expression (\ref{jane.effpower}). The result, for the initial condition (\ref{jane.cosh}) is shown in Fig.~2. 

\begin{figure}
\label{janfig2}
\begin{center}
\includegraphics[height=4cm]{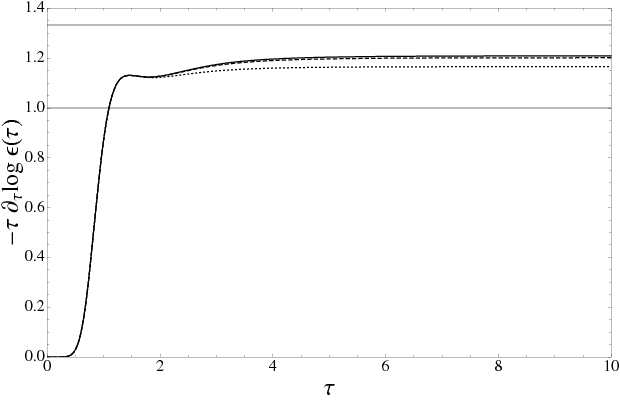}
\end{center}
\caption{The effective power (\ref{jane.effpower}) of $\eps(\tau)$ corresponding to the initial condition (\ref{jane.cosh}).}
\end{figure}

We see that it definitely crosses $s=1$ of free streaming and moves upwards. However, to be sure that it would reach $4/3$ a numerical solution for 
$\eps(\tau)$ would be needed.

Now assuming the late time exponent $4/3$, we may perform a Pade approximation of $\eps(\tau)$ with this asymptotics to see the profiles of $\eps(\tau)$ for a set of initial conditions. Example plots are shown in Fig.~3.

\begin{figure}
\label{janfig3}
\begin{center}
\includegraphics[height=2cm]{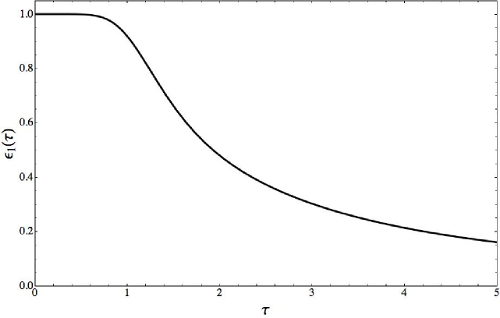}
\hspace{0.5cm}
\includegraphics[height=2cm]{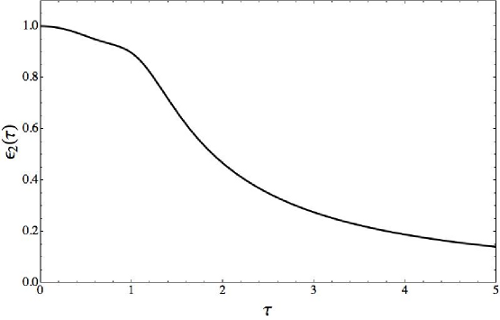}
\hspace{0.5cm}
\includegraphics[height=2cm]{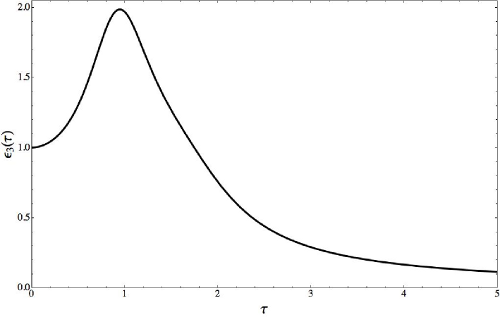}
\end{center}
\caption{Pade resummed profiles of $\eps(\tau)$ for a set of initial conditions \cite{janusearly}.}
\end{figure}

The main interest of the knowledge of the exact profile of $\eps(\tau)$ for various initial conditions is that then we would be able to study the  transition to hydrodynamics and its dependence on various features of the initial conditions.  Since the Pade extrapolation introduces some serious systematic uncertainties, we refrain from doing so until we will have at our disposal a direct numerical solution of Einstein equations for these various initial conditions. Apart from the reasons mentioned above, the numerical solution would also allow to analyze the nature of the apparent singularity in the initial data and, more importantly, analyze the geometry for the presence of apparent (dynamical) horizons, relevant for the thermodynamic interpretation. Some of these issues are currently under investigation \cite{janWIP}.

Before we close this last part of the lectures let us note a complementary numerical investigation of boost invariant plasma in \cite{janboostcy}. The setup of \cite{janboostcy} was different from the one presented in these lectures in that the \emph{gauge theory metric} was perturbed in a boost invariant way at some $\tau \sim \tau_0>0$ and then set again to flat Minkowski. The metric perturbation produced a change in the geometry which induced a boost invariant flow on the boundary. The main results observed in \cite{janboostcy} were a transition to hydrodynamics and a formation of an apparent horizon which moved in from infinity.

Another related work\footnote{This time not in the boost-invariant setting discussed here.} was \cite{janminwallabh}, were a perturbation by a boundary scalar source induced a black hole formation in the bulk.

Looking at all these examples, one sees that Einstein's equations, through the AdS/CFT correspondence, have the potential of describing a multitude of far from equilibrium strongly coupled gauge theory phenomenae. It is however clear that the majority of problems remain still unsolved.

\section{Conclusions}

In these lectures we have described an approach using the AdS/CFT correspondence as a tool for studying real time dynamics of strongly coupled gauge theory plasma. The basic theoretical tool is the possibility of translating, in a completely explicit and constructive way, between the spacetime energy-momentum tensor characterizing the gauge theory configuration in question and the five-dimensional metric of the dual geometry. Then one uses the fact, that at strong coupling, the dynamics of the dual geometry is governed by Einstein's equations (with a cosmological constant following from the full 10 dimensional supergravity solution). A further dynamical input is the requirement of the absence of naked singularities in the gravity background. This restricts very strongly the allowed spacetime profiles of the gauge theory energy momentum tensor, and consequently the possible gauge theory dynamics.

Using these methods, one may establish the appearance of nonlinear hydrodynamics, as exhibited here in the form of near perfect fluid dynamics in the large proper time asymptotics of boost invariant plasma expansion. Moreover, Einstein equations together with the nonsingularity criterion unambigously predict viscous first- and higher-order deviations from perfect fluid dynamics with specific values of the transport coefficients appropriate to the case of the $\nn=4$ SYM theory studied here. 

Let us note, that we arrived at these results without presupposing anything even about the general form of the gauge theory energy momentum tensor like the presence of something like a flow velocity $u^\mu$ etc. 
This flexibility comes from the fact that Einstein's equations on the string side of the AdS/CFT correspondence require only that the gauge theory energy-momentum tensor is conserved and traceless (throughout these lectures we are dealing exclusively with the conformal case of $\nn=4$ SYM). Therefore one can use the same techniques to address the question of far from equilibrium dynamics where hydrodynamics cannot be used as a starting point of approximation. We exhibited an example of such a study by describing some aspects of the behaviour of the boost invariant plasma expansion in the region of small proper time.

It should be obvious that the majority of questions concerning far from equilibrium dynamics of strongly coupled plasma remain still unanswered even for the case of $\nn=4$ plasma. Once more details are understood, it would be very interesting to address similar problems in more complicated versions of the AdS/CFT correspondence involving gauge theories which might be closer to QCD.
Apart from this direct `application driven' motivation, the study of such time-dependent systems leads to fascinating interrelations with General Relativity. On the one hand, it provides a new setting for investigating such GR concepts as dynamical apparent horizons, black hole formation, providing these GR phenomenae with new interpretation. On the other hand, the well developed technology of numerical relativity might be applied to learn more about the properties of far from equilibrium strongly coupled gauge theory systems.

Last but not least, let us note that the gravity backgrounds obtained as dual descriptions of evolving plasma systems can very well by themselves form a scene for the AdS/CFT correspondence enabling one to study various kind of physics questions. In this way one may study how the expanding plasma system influences properties of mesons, Wilson loops, various correlation functions. Of course, due to the time-dependent nature of those backgrounds this may be quite difficult to do in practice, but the possibility of doing so is certainly very interesting and the results may be rewarding.

\begin{acknowledgement}
The approach presented in these lectures was introduced with Robi Peschanski, and developed further in collaboration with Michał Heller, Dongsu Bak, Alex Buchel, Paolo Benincasa, Guillaume Beuf. To all of whom I am grateful for enjoyable collaboration and numerous discussions. I would like to thank the organizers and participants of the \emph{Fifth Aegean Summer School "From Gravity to Thermal Gauge Theory: The AdS/CFT Correspondence"} for a very interesting school. I would also like to thank the \emph{New Frontiers in QCD 2010} program and the Yukawa Institute for Theoretical Physics, Kyoto for hospitality when these lectures were written up. This work was supported in part by Polish science funds as a research project N N202 105136 (2009-2011) and the Marie Curie ToK KraGeoMP (SPB 189/6.PRUE/2007/7). 
\end{acknowledgement}

\appendix

\section*{Appendix}

\subsubsection*{Topics not covered in the main text --- a brief guide to the literature}

In this appendix we would like to give pointers to the literature on other developments related to the approach presented in these lectures.

\begin{itemize}
\item Leading $\al'$ corrections (i.e. beyond strong coupling in $\nn=4$ SYM theory)  to the transport coefficients were computed using the boost invariant flow \cite{janr72,janr53}. This task involved using $\al'$ corrected Einstein's equations.

\item Beyond $\nn=4$ SYM. A class of general conformal field theories parametrized by higher curvature terms in the dual gravitational action was considered in \cite{janr7}.

\item Extension to hydrodynamics with conserved charge(s) was considered in \cite{janr99,jansur,janr23}.
Electric-magnetic equilibration at large proper times was found in \cite{janr99}. In~addition, dilaton driven hydrodynamics was considered in a general way 
in~\cite{janr62} and the onset of turbulence was observed \cite{janminwallaturb}.

\item Lower dimensional examples. The case of a 1+1 dimensional conformal field theory allows for an explicit exact dual gravitational solutions 
\cite{janr93,janr74}. Other investigations in different number of dimensions were performed in \cite{janr50}.

\item Various (exact) solutions for $\nn=4$ gauge theory in curved and possibly time-dependent backgrounds were obtained \cite{janr105,janr98,janr49}. The exact solutions which were found do not involve viscosity effects. An exact gravitational description af a shearless flow in $\nn=4$ in flat space was obtained in \cite{janr102}.

\item Further properties of solutions with boost invariant symmetries were studied in~\cite{janr111,janr78,janr48,janr40,janr16}.

\item Physics in the expanding plasma. Fundamental flavours were introduced (through D7 brane embeddings) into the late proper time boost invariant geometry \cite{janr84}, diffusion constant was evaluated \cite{janr89}, drag force was computed \cite{janr22}. In addition, various physical questions were addressed in the case of the shock wave solutions described in section~\ref{jans4}.
Deep Inelastic Scattering (DIS) was analyzed \cite{janr60,janr44,janr12,janr2} as well as heavy quark energy loss \cite{janr21,janr10}.

\end{itemize}

\end{document}